\documentclass[12pt]{article}
\usepackage{pic03}
\usepackage{hyperref}
\usepackage{url}
\usepackage{graphicx}

\begin{document}

\title{\bf 
  MONTE-CARLO SIMULATION OF EXCLUSIVE CHANNELS
  IN $e^+e^-$ ANNIHILATION AT LOW ENERGY}

\author{
  Dmitry Anipko \\
  Simon Eidelman \\
  Alexey Pak \\
  {\em Novosibirsk State University, Pirogova st. 2, 
       Novosibirsk 630090, Russia}\\
  {\em Budker Institute of Nuclear Physics, Lavrentjeva pr. 11, 
       Novosibirsk 630090, Russia}
}

\maketitle
\baselineskip=14.5pt

\begin{abstract}
  Software package for Monte-Carlo simulation of $e^+e^-$ exclusive 
  annihilation channels written in the C++ language for Linux/Solaris 
  platforms has been developed. It incorporates matrix 
  elements for several mechanisms of multipion production in a model 
  of consequent two and three-body resonance decays.
  Possible charge states of intermediate and final particles are 
  accounted automatically under the assumption of isospin conservation. 
  Interference effects can be taken into acccount. Package structure allows 
  adding new matrix elements written in a gauge-invariant form.
\end{abstract}

\baselineskip=17pt

\section{Introduction}
  Simulation of hadron production at low energies ($\sqrt{s} \sim 1-3$ GeV) 
  is relevant for various physical problems. 
  As the cross-section in this region is considered to be saturated 
  with intermediate resonances, it is possible to study the properties 
  of these resonances through mass, momentum and angular distributions. 
  Simulation of different channels allows one to better account for 
  selection rules and interference effects. In particular, the problem
  of measuring R (total cross-section of $e^+e^-$ annihilation into 
  hadrons) is a crucial information for contemporary high energy physics,
  especially for the $\frac{g_\mu-2}{2}$ problem \cite{muonG2}. At 
  low energies this value can only be obtained from an experiment. 
  Simulation allows to calculate the ratios between different channels 
  and estimate the accuracy of measurements.

  Of other related problems we can highlight the studies of 
  $e^+e^-\to6\pi$ at $\sqrt{s} < m_{\tau}$ and $\tau^- \to (6\pi)^- \nu_{\tau}$ decays
  allowing to check the vector current conservation hypothesis in the Standard Model 
  \cite{CVC6PI,CLEOTAU6PI}, studies of the $\rho^{\prime\prime}\to 6\pi$ decay 
  mode as well as $D$- and $B$-meson decays.

\section{Package features}
  At this moment there is no theory, describing the strong 
  interactions reliably at low energies. We used a common
  phenomenological approach, assuming the production of a 
  final state in consequent resonance decays: \\
  $\bullet$ only tree-level diagrams including 2- and 3-body decays, 
          are considered; \\
  $\bullet$ there may be several interfering mechanisms 
         (e.g. $\gamma^\ast\to\pi^\pm a_1^\mp,\rho_0 \sigma$)
         which may contribute to different final states 
         (e.g. $2\pi^+2\pi^-$,$\pi^+\pi^-2\pi^0$) in accordance 
         with charge and strong isospin conservation; \\
  $\bullet$ possible permutations of final particles are taken into account.\\
  To simplify adding new matrix elements, we introduced the following 
  requirements: \\
  $\bullet$ the tensor structures are chosen according to the 
         field transformation properties (spin, isospin, parities);\\
  $\bullet$ expressions for matrix elements are written in gauge-invariant
         form with the help of specialized tensor library
         (e.g., 
         $\epsilon_{\xi \theta \sigma \tau} P_{\rho_{init}}^\sigma 
         \epsilon_{\xi \theta \vartheta \delta} P_{a_1 \vartheta} 
         \epsilon_{\mu \nu \gamma \delta} P^{a_1 \gamma}  
         \epsilon_{\mu \nu \alpha \beta} 
         P_{\rho_{12}\alpha} (P_{\pi_1}-P_{\pi_2})_{\beta}$); \\
  $\bullet$ indices' contraction and particles permutations are 
         performed by the software.\\ 
  Allowed charge states and relative phases of permutations are 
  based on the strong isospin part of the matrix element.
  Scalar parts of propagators may have arbitrary forms, including 
  form-factors and dependence of width on virtuality.
  In case of the vector intial state ($\gamma^\ast$), the absolute 
  value of the transverse part of hadronic current is used as the 
  value of matrix element. Several interfering matrix elements 
  with arbitrary complex relative coefficients may be used.


\section{Comparison with experimental data}
\begin{figure}
    \hspace{0.1cm}
    \includegraphics[width=6.9cm,height=160pt]{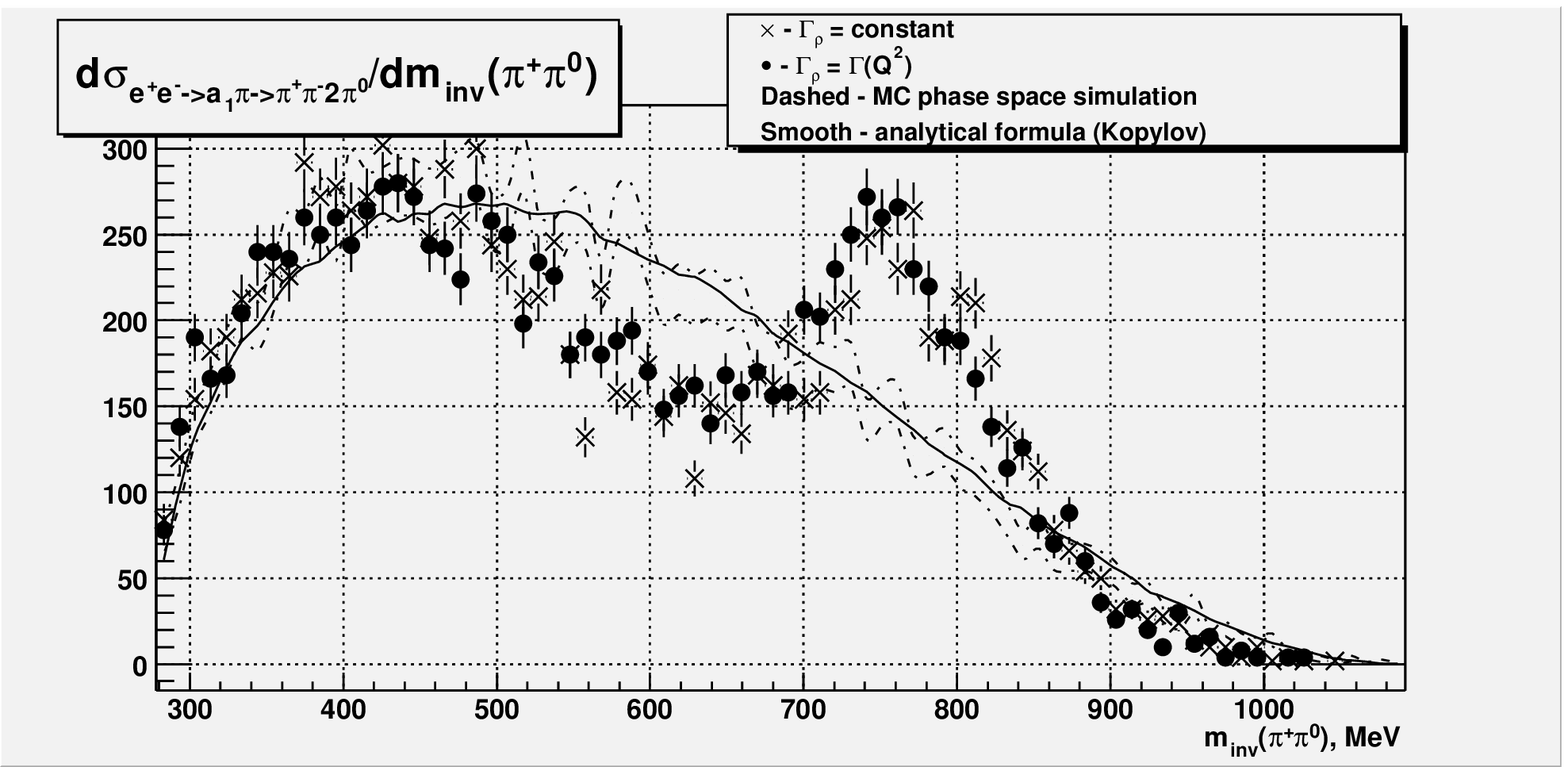}
    \hspace{0.1cm}
    \includegraphics[width=6.9cm,height=160pt]{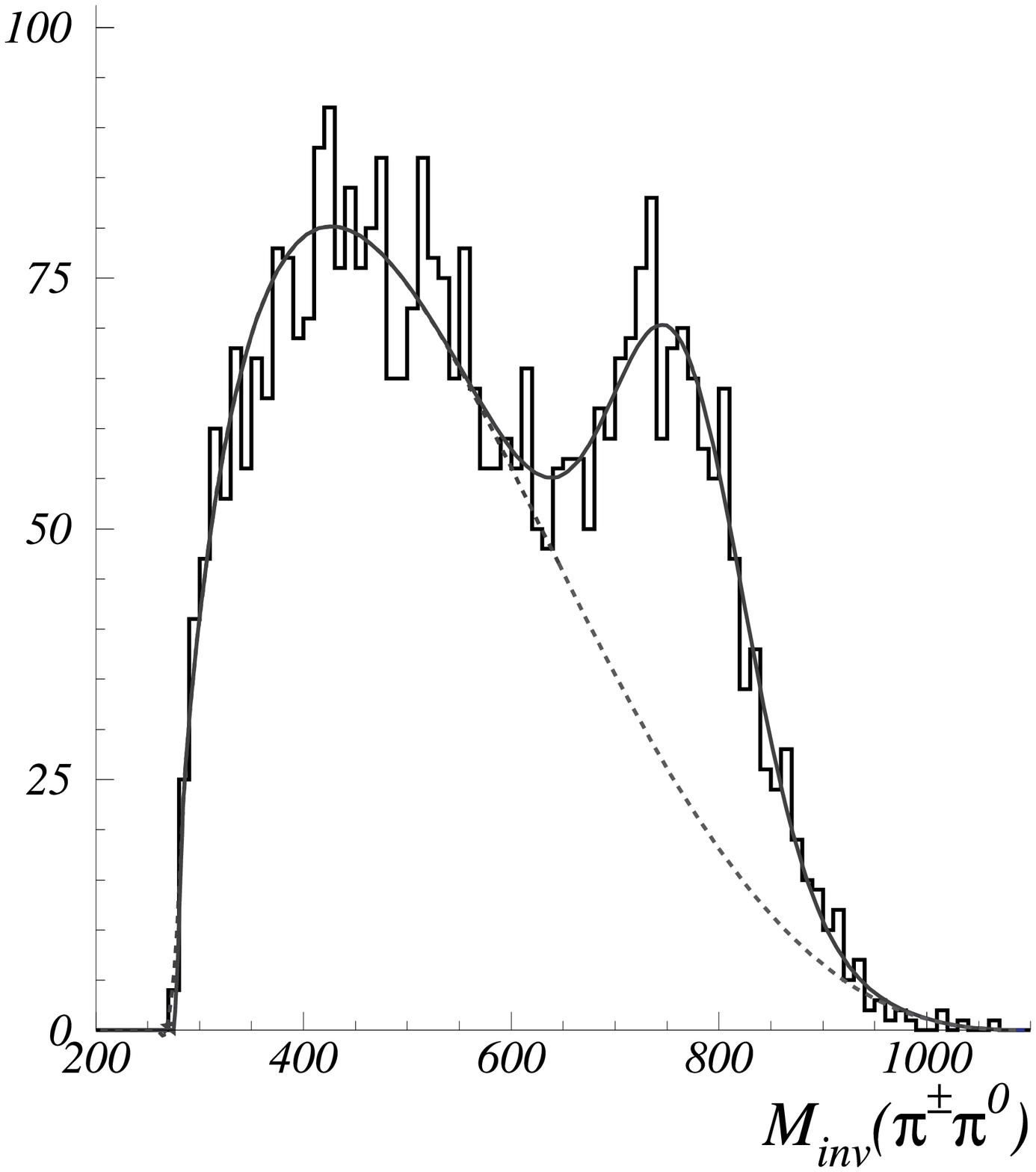}
    \caption{Distribution on dipion invariant mass $m_{\pi^+\pi^0}$ 
             in the channel $e^+e^-\to a_1\pi\to\rho\pi\pi\to\pi^+\pi^-2\pi^0$ 
             -- our simulation (left) and experimental data (right).}
    \label{2picomp}
  \end{figure}

  By now the package has been used to calculate 4 and 6-pion
  production cross-sections in $e^+e^-$-collisions. Some distributions
  of simulated data vs. experimental results \cite{CMD4PI} are given 
  in Fig. \ref{2picomp} (parameters not fitted).
  Comparison with analytical calculations given in Ref. \cite{Kopylov}
  show a reasonable agreement.\\
  At the VEPP-2000 collider (the major upgrade of VEPP-2M, BINP, 
  Novosibirsk) $6\pi$ production will become possible. 
  Despite the lack of absolute normalization factor in matrix 
  elements, the ratio of cross-sections of different final 
  charge states with   the same production mechanism can be 
  tested for $6\pi$ data.

\section{Conclusion}
  The package for simulation of multipion production has 
  been developed. It includes generators for
  $\gamma^\ast \Rightarrow$ $\rho\pi \to 3\pi$,
  $a_1\pi, \rho f_0 \to4\pi$,
  $\omega\pi\pi\to5\pi$, and
  $\omega\pi, \rho f_0 (\rho\rho, \sigma\sigma) \to 6\pi$
  production mechanisms. This tool may be useful for many 
  problems involving production of multipion final states.

\section{Acknowledgements}
  We are thankful to O.Yu.~Dashevskij, I.F.~Ginzburg, P.P.~Krokovny, A.S.~Kuzmin, A.I.~Milstein
  and N.I.~Root for useful discussions
  and helpful contributions. D.A. thanks DFG foundation for support of 
  his participation in PIC2003. The work is supported by
  RFBR grants 02-02-17884-a and 03-02-06651-mac.

\end{document}